\documentclass[final,5p,times,twocolumn]{elsarticle}

\usepackage{graphicx}
\usepackage{amssymb}
\usepackage{amsmath}
\usepackage{multirow}
\usepackage{color}

\newcommand{\action}{\mathbf{a}}

\journal{Elsevier Computer Communications 2 July 2020, Received in revised form 22 November 2020, Accepted 7 January 2021}

\begin{document}

\begin{frontmatter}


\title{Increasing Energy Efficiency of Massive-MIMO Network via Base Stations Switching using Reinforcement Learning and Radio Environment Maps
}



\author{Marcin Hoffmann}
\ead{marcin.ro.hoffmann@doctorate.put.poznan.pl}

\author{Pawel Kryszkiewicz}
\ead{pawel.kryszkiewicz@put.poznan.pl}

\author{Adrian Kliks}
\ead{adrian.kliks@put.poznan.pl}
\address{Institute of Radiocommunications, Poznań University of Technology, Poznań, Poland}

\begin{abstract}

Energy Efficiency (EE) is of high importance while considering  Massive Multiple-Input Multiple-Output (M-MIMO) networks where base stations (BSs) are equipped with an antenna array composed of up to hundreds of elements. M-MIMO transmission, although highly spectrally efficient, results in high energy consumption growing with the number of antennas.  
This paper investigates EE improvement through switching on/off underutilized BSs. It is proposed to use the location-aware approach, where data about an optimal active BSs set is stored in a Radio Environment Map (REM). For efficient acquisition, processing and utilization of the REM data, reinforcement learning (RL) algorithms are used. State-of-the-art exploration/exploitation methods including $\epsilon$-greedy, Upper Confidence Bound (UCB), and Gradient Bandit are evaluated. Then analytical action filtering, and an REM-based Exploration Algorithm (REM-EA) are proposed to improve the RL convergence time. Algorithms are evaluated using an advanced, system-level simulator of an M-MIMO Heterogeneous Network (HetNet) utilizing an accurate 3D-ray-tracing radio channel model. The proposed RL-based BSs switching algorithm is proven to provide 70\% gains in EE over a state-of-the-art algorithm using an analytical heuristic. Moreover, the proposed action filtering and REM-EA can reduce RL convergence time in relation to the best-performing state-of-the-art exploration method by 60\% and 83\%, respectively.         
\end{abstract}

\begin{keyword}
Massive MIMO \sep Radio Environment Map \sep Base Station Switching On/Off \sep Reinforcement Learning \sep Energy Efficiency


\end{keyword}

\end{frontmatter}


\section{Introduction}





In the last decades we could observe a huge growth in the deployment of mobile telecommunication networks. It is caused by an increasing number of mobile devices: smartphones, tablets, but also Internet-of-Things (IoT) devices. With the increased number of mobile devices, and rapidly growing network infrastructure, energy-efficiency (EE) became a major field of research aiming at reducing operators' costs and carbon emissions~\cite{hasan2011}. EE becomes even more challenging, when considering the 5G Massive Multiple-Input Multiple-Output (M-MIMO) network utilizing large antenna arrays~\cite{larsson2014}. Although M-MIMO enables reduction of the transmitted power, through high effective antenna gain obtained using beamforming, it is not always enough to compensate the power consumed by hardware components e.g., transceivers chains, filters, modulators~\cite{massivemimobook}. Moreover, the 5G system will be typically a heterogeneous network (HetNet), i.e., a  network composed of overlapping cells of various sizes, with each Base Station (BS) equipped with an antenna array \cite{prasad2017}. Although HetNet can provide large spectral-efficiency benefits, increased energy consumption is expected.

The energy consumption decrease is acceptable only if the Quality of Service (QoS) for the end users is not degraded significantly.
However, it has been shown that traffic load in a telecommunication network varies over time \cite{jin2012characterizing, shafiq2011characterizing}. As such, many of the BSs are underutilized within some time periods. There are several ways of adjusting BSs configuration to this phenomena and reducing the energy consumption. First, the antennas and related transceiver chains receiving the least amount of power can be switched off \cite{gao2015}. Secondly, some energy savings can be provided by reducing the resolution of analog-to-digital converters, by up to one bit \cite{mollen2017}. Thirdly, energy consumption can be decreased, when hybrid analog-digital precoding is used~\cite{zi2016}. But the highest reduction of the consumed power can be achieved by switching off underutilized BSs. 

There are several base station on/off switching algorithms proposed in the literature \cite{han2016, wu2015}. Some of them obtain a subset of active BSs through optimization methods. Others propose analytical heuristics, considering the traffic load, or a number of User Equipments (UEs) served by each BS. Another approach is to observe the performance of a chosen active BSs set, gather knowledge and make an \emph{a posteriori} decision about the optimal active BSs set. This can be achieved by utilizing a subclass of machine learning (ML) algorithms, called reinforcement learning (RL) \cite{sutton2018rl}. There are a few works where RL is considered for providing energy savings in wireless networks \cite{sharma2019rl,islam2019rl, ye2019drag}.  
The idea of the existing RL-based BS-switching algorithms relies mostly on the utilization of explicit traffic load or user association information.

However, in an M-MIMO network simple network models do not suffice. High-level information about UE influence on a system's EE can not be accurate enough. This is the result of significantly varying channel conditions depending on UE location. Fortunately, 5G comes with improved localization techniques \cite{hoffmann2020}, enabling the so-called location-aware communications \cite{taranto2014}. Accurate localization allows to tag the measurement of various radio link parameters with UE coordinates. Such location-dependent information can be stored in the so-called Radio Environment Map (REM) \cite{tengvist2017}. REM is an intelligent unit attached to the existing network infrastructure responsible for the acquisition, storage, and processing of location-dependent information with the aim to improve network performance. 

In this paper, the 5G M-MIMO network is subject to EE maximization while keeping high QoS for the users. Although in the early stages of 5G rather hybrid antenna arrays, i.e., a limited number of digital transceiver chains combined with an analog processing network, would be deployed in order to reduce the implementation cost, we are considering full digital architecture. This is proper for fully developed 5G networks pointing toward 6G~\cite{6gwp2020, guo2019}. The EE can be changed by the activation/deactivation of BSs. The considered network is highly complex, including an advanced radio channel model (3D ray tracing-based), realistic channel estimation inducing errors, user-to-BS association, scheduling, and precoding algorithms. This precludes simple, yet accurate problem formulation in order to utilize standard optimization methods. Instead, Reinforcement Learning (RL) has been used, where the active BS set can be changed to observe \emph{a posteriori} changes in M-MIMO network Quality of Service (QoS) metrics. The high influence of UEs location on M-MIMO network performance is exploited by the utilization of REM for storing RL states. This allows the network to preserve the learned active BSs set for each UE distribution pattern. In order to increase the convergence speed of the proposed algorithm (as in a real network, BS deactivation is a serious operation possibly disconnecting some UEs from the network), analytical possible action filtering is implemented, i.e., a reduction of the possible set of active BSs to be considered in RL. In addition, an information fusion function is proposed in order to utilize in a given network state (i.e., for a certain UEs location pattern) information about past network performance in the adjacent states, i.e., REM-based Exploration Algorithm (REM-EA). The \emph{distance} between the states will be specified using Haussdorf distance. This helps to improve learning convergence even further, while not deteriorating the final EE of the system.

The paper is organized as follows: In Sec.~\ref{sec:related_work} an overview of the related works is presented. The considered system is described in Sec.~\ref{sec:system_description}, while Sec~\ref{sec:rem} presents an idea of REM, and describes REM deployment in the considered M-MIMO HetNet. The RL concept adopted to BS switching in M-MIMO HetNet and the proposed REM-based algorithms are described in Sec.~\ref{sec:rl}. The simulation environment and results are presented in Sec.~\ref{sec:simulation}. Finally, the conclusions are formulated in Sec.~\ref{sec:conclusions}.

\section{Related Work} \label{sec:related_work}

A vast number of algorithms providing energy savings by switching off underutilized BSs can be found in the literature. A major group of the proposed solutions comprise algorithms based on simplified network models and optimization heuristics \cite{han2016}. One of the prospective ideas is to turn off pico or femto BSs having no or just a few UEs in their coverage area to serve \cite{liu2016}. A more advanced approach relies on switching off BSs on the basis of a traffic load threshold \cite{Abdulkafi2014}. Another solution, is based on switching off a particular BS only if another BS has enough available radio resources to take over the additional load \cite{oh2013swes}. However, the authors of these articles did not consider M-MIMO BSs, and utilized simple radio channel models. Thus, the proposed analytical heuristics are inadequate to be implemented under the real radio conditions. The authors of \cite{feng2017boost} proposed a distributed solution for energy-efficient BS switching based on the game theory. Although there is M-MIMO considered, the antenna array is claimed to be deployed only at the macro BS (MBS), while small BSs are equipped with single antennas only. The utilized spectral efficiency estimate for MBS is simplified, i.e, it neglects inter-UE interference, and assumes that every UE can be served using an independent spatial beam. Finally, not very accurate radio channel model is assumed. It consists only of the pathloss and Rayleigh fading, not considering, e.g., spatial correlation affecting M-MIMO performance. 

On the other hand, there are several algorithms utilizing reinforcement learning for providing an EE gain through BS switching. In \cite{sharma2019rl}, there is an actor-critic algorithm utilized to switch off BSs, so as to minimize the overall system power consumption. The action selection is done on the basis of a stochastic policy following Boltzman distribution, while environment state is defined as the total traffic load per BS. A more advanced solution takes into account not only power consumption, but also latency, and the number of served UEs \cite{wang2014rl}. Additionally, the authors utilize Boltzman distribution-based action selection, though with the so-called Q-learning algorithm. The authors of \cite{islam2019rl} proposed a three-state (i.e., active, stand-by, and sleep) Markov Decision Process for BS, and a Value Iteration RL framework to minimize network power consumption. Also there is the BS switching algorithm, utilizing the so-called Deep Reinforcement Learning, i.e., RL learning boosted with the use of a deep neural network~\cite{ye2019drag}. The cost function including power consumption, QoS, and switching cost is minimized on the basis of the input traffic load estimates and active BSs set. 

The major issue of all mentioned RL algorithms is the lack of M-MIMO consideration in the system model. Secondly, similarly to the analytical heuristic methods, the RL algorithms are evaluated using simple radio channel models, e.g., considering only the pathloss component and Rayleigh fading. Although algorithms were accurately studied in terms of their final performance, less effort was put into the evaluation of their convergence time. Fast convergence is crucial not only for minimizing BS hardware expenditure, but also for reducing QoS degradation related to the learning process, i.e., trying actions that degrade QoS to obtain \emph{a posteriori} knowledge. 

In this paper, we propose to utilize the knowledge about UE location in the M-MIMO network instead of explicit traffic load information used in the recent works. We will show that this approach can simplify the full RL problem to the so-called \emph{Associative Search}. In contrast to the majority of the mentioned works, in this paper the convergence of state-of-the-art RL exploration methods will be extensively studied. In addition, an analytical action filtering algorithm is proposed to speed up RL convergence. Moreover, the proposed REM-EA utilizes the similarities between UEs position sets, using data stored in REM, to further improve RL convergence. The gains of the proposed algorithms over the existing exploration strategies are proven through computer simulations. In comparison to the other recent works, in this paper, an advanced system level simulator is implemented that utilizes a realistic 3D-ray-tracing radio channel model. Following the studies were models considering regular deployment of BSs as they were claimed as inaccurate in practice, we decided to evaluate the proposed algorithms using the realistic \emph{Madrid Grid Model}, used previously in, e.g., METIS project~\cite{metis2020}. It would be beneficial to study random BS locations, as proposed in many contemporary works, e.g.,~\cite{ge2015}, but the complexity of the simulated system prevents us from such considerations. Maybe this would be possible in the future for a simplified system model. 

\section{System Description} \label{sec:system_description}
\begin{figure}[h]
\centering\includegraphics[scale=0.32]{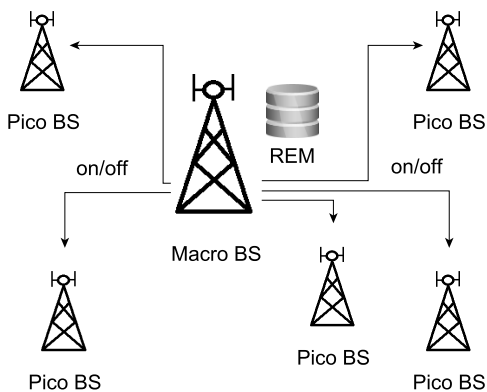}
\caption{Scheme of the considered M-MIMO HetNet}
\label{fig:cell}
\end{figure}

In this paper we consider switching BSs on/off in a single, two-tier HetNet. The network consists of a MBS, and several pico BSs (PBSs). All BSs in the system share the same radio frequency resources, and utilize Orthogonal Frequency Division Multiple Access (OFDM). All BSs in the HetNet are equipped with large antenna arrays exploiting from tens up to hundreds of antenna elements. The number of antenna elements, and array design can vary between the consecutive BSs. We assume that the MBS is not considered for switching off as it provides a general coverage in the cell and performs the centralized management functions. Commonly, the utilized radio access technologies have a fixed set of Modulation-Coding Schemes (MCS) with the associated minimum Received Signal Strength (RSS) that allows then to serve a given UE. This approach is followed in this paper. The process of switching BSs on/off is also centralized and managed from the MBS, where REM and RL algorithms are deployed, as depicted in Fig.~\ref{fig:cell}. 

The REM deployed in the Macro BS has access to all control information available in the BSs, e.g., UE bitrates, UE to BS association, channel estimates, power consumption. In this paper, downlink bitrates will be of interest. This information is then utilized to formulate REM entries necessary for the RL base station switching algorithms.

\subsection{Power Modeling} \label{subsec: power_model}
There are several BS power consumption models proposed for the LTE system~\cite{holtkamp2013, capone2017}. Unfortunately they are not adequate for M-MIMO evaluation. M-MIMO utilizes tens or hundreds of antennas per BS, which is much more than the maximum of 8 antenna elements available in LTE-A~\cite{DBLP:journals/ejwcn/LeeHZ09}. Therefore, in this work, the model for the evaluation of M-MIMO BS energy consumption proposed in \cite{massivemimobook} is used. Mainly, the power consumption of all $N_{\mathrm{BS}}$ BSs  is composed of 3 components: effective transmitted power (ETP), transceivers chains power, and fixed power. ETP stands for the energy consumption related to the total radiated power $P_{\mathrm{tx},b}$ by BS $b$ and power amplifier efficiency $\eta$. It is given by:
    \begin{equation}
        \label{eq:etp}
        P_{\mathrm{ETP},b} = \frac{P_{\mathrm{tx},b}}{\eta}.
    \end{equation}
The transceivers chain power in the $b$-th BS is denoted as $P_{\mathrm{tc},b}$. It describes the power consumed by the hardware utilized by every single antenna (such as modulators, filters, etc.) and a local oscillator (LO). It can be computed as:
        \begin{equation}
            \label{eq:cp_transciever}
            P_{\mathrm{tc},b} = M_b P_{\mathrm{\widehat{TC}}} + P_{\mathrm{LO}},
        \end{equation}
where $M_b$ stands for the number of antennas deployed at BS $b$, $P_{\mathrm{\widehat{TC}}}$ models the amount of power consumed by the single transceiver chain, and $P_{\mathrm{LO}}$ is the power necessary for the LO to work. 

The last component, fixed power $P_{\mathrm{fix}}$ denotes the power necessary for, e.g., baseband signal processing, in each BS.

In this paper, by switching off the BSs we put them in the stand-by-mode, where the utilization of hardware resources is much reduced. In consequence, a BS operating in the stand-by-mode consumes in total power $P_{\mathrm{off}}$. Moreover, it is assumed that such a BS can be switched on rapidly, e.g., within $30$ $\mu s$ \cite{frenger2011}. Such a short switching time enables immediate changes in the active BSs set. Deeper sleep/switch off, although more energy efficient, will require much more time for transition between on/off states, and due to that, it is not considered in our paper.  

The overall system power consumption $P_{\mathrm{tot}}$ is given by:
\begin{equation}\label{eq:p_tot}
    P_{\mathrm{tot}} = \sum_{b=1}^{N_{\mathrm{BS}}}P_{\mathrm{tot},b},
\end{equation}
where
\begin{equation} \label{eq:p_tot2}
P_{\mathrm{tot},b} = \begin{cases} P_{\mathrm{ETP},b} +  P_{\mathrm{fix}} + P_{\mathrm{tc},b}, & \text{for active BS} \\ P_{\mathrm{off}}, & \text{for BS in stand-by-mode} 
\end{cases}
\end{equation}


\section{Radio Environment Map} \label{sec:rem}

The idea of acquisition of location-aware data in order to improve network performance comes from cognitive radio~\cite{sodagari2015}. There is a concept of the Radio Environment Map (REM) proposed to gather location-dependent information about the surrounding radio environment to enable opportunistic transmission for secondary users~\cite{kliks2017}. Although REM generally stores interference power or received signal strength values, the idea can be easily extended to include any context information that can be useful for network operation. As a result, REMs can store location-aware data about e.g., UEs velocity or network traffic load, and as such are sometimes called Radio Service Maps~\cite{tengvist2017}. In this paper, REM stores information processed by the learning procedure that include UEs positions and achievable EE depending on the active BSs set. 

\subsection{REM Deployment}
\begin{figure}[h]
\centering\includegraphics[scale=0.32]{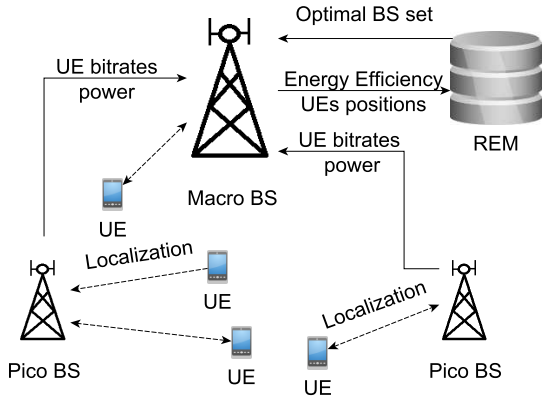}
\caption{Scheme of REM deployment in the considered M-MIMO HetNet. 
}
\label{fig:rsm_in_cell}
\end{figure}

The deployment of REM in the considered M-MIMO HetNet is depicted in Fig.~\ref{fig:rsm_in_cell}. We assume that a so-called REM agent is installed on the MBS. The first objective of the REM agent is to process the data provided by the UEs and BSs, and store them in REM. Secondly, the REM agent optimizes the current active BSs set according to the proper algorithm, and using REM data. UEs provide their locations measured with state-of-the-art localization methods, e.g., using highly accurate satellite navigation \cite{fortunato2019}. Each BS calculates the associated UEs bitrates and its power consumption. The REM agent calculates the network EE metric based on these inputs and stores it characterized by the given UEs position set. On the basis of the previous measurements, and according to the chosen learning algorithm, the REM agent chooses the next active BSs set.  

\subsection{REM Data Structure} \label{subsec:rem_data_structure}
\begin{figure}[h]
\centering\includegraphics[scale=0.37]{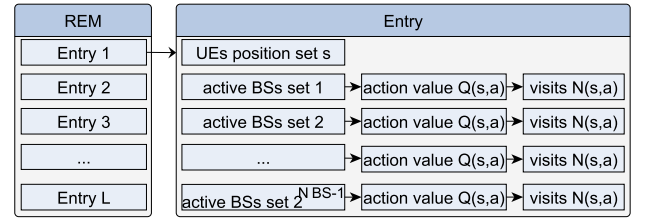}
\caption{Data structure in REM. While MBS is not considered for switching off, there are $2^{N_{\mathrm{BS}}-1}$ possible active BSs sets.  
}
\label{fig:bs_switching_rsm}
\end{figure}

The structure of an REM database is presented in Fig.~\ref{fig:bs_switching_rsm}. Each entry is identified on the basis of the UEs positions set $\mathbf{s}$, being in fact a vector of the UEs coordinates on the Cartesian plane, rounded to fit a square grid. The length of the considered grid square side is denoted as  $g$. 
In order to evaluate the similarity between different UEs position sets $\mathbf{s}$, the so-called Haussdorf distance (HD) metric is used \cite{huttenlocher1993}. The main reason to choose HD is its ability to compare data sets hadding a different number of measurements, i.e., UEs in our case. Suppose there have been $N_{\mathrm{UE}}$ UEs present in the network at positions $\mathbf{s_{\mathrm{REM}}} = \{s_{\mathrm{REM},1}, s_{\mathrm{REM},2}, \hdots, s_{\mathrm{REM},N_{\mathrm{UE}}}  \}$ with $s_{\mathrm{REM},n}$ being the position of $n$-th UE. The HD to the current position set of $N_{\mathrm{UE}^\prime}$ UEs $\mathbf{s_{\mathrm{0}}} = \{s_{\mathrm{0},1}, s_{\mathrm{0},2}, \hdots, s_{\mathrm{0},N_{\mathrm{UE}^\prime}}  \}$ is given by \cite{SIM2016523}:
\begin{equation} \label{eq:hd}
    d_h(\mathbf{s_{\mathrm{REM}}}, \mathbf{s_{\mathrm{0}}}) = \max (hd(\mathbf{s_{\mathrm{REM}}}, \mathbf{s_{\mathrm{0}}}), hd(\mathbf{s_{\mathrm{0}}}, \mathbf{s_{\mathrm{REM}}})),
\end{equation}
\begin{equation}
    \label{eq:hd2}
    hd(\mathbf{s_{\mathrm{REM}}}, \mathbf{s_{\mathrm{0}}}) = \max_{s_{\mathrm{REM}} \in \mathbf{s_{\mathrm{REM}}}} \{ \min_{s_{\mathrm{0}} \in \mathbf{s_{\mathrm{0}}}} (d_e (s_{\mathrm{REM}},s_{\mathrm{0}} ) ) \},
\end{equation}
\begin{equation}
    \label{eq:hd3}
    hd(\mathbf{\mathbf{s_{\mathrm{0}}}, s_{\mathrm{REM}}}) = \max_{s_{\mathrm{0}} \in \mathbf{s_{\mathrm{0}}}} \{ \min_{s_{\mathrm{REM}} \in \mathbf{s_{\mathrm{REM}}}} (d_e (s_{\mathrm{0}},s_{\mathrm{REM}} ) ) \},
\end{equation}
where $d_e(\cdot, \cdot)$ stands for Euclidean distance. On the basis of $d_h(\mathbf{s_{\mathrm{REM}}}, \mathbf{s_{\mathrm{0}}})$, a decision is made if the new UEs position set is recognized as one of the REM entries or not. The UEs position set is recognized as an existing REM entry if  $d_h(\mathbf{s_{\mathrm{REM}}}, \mathbf{s_{\mathrm{0}}}) < g$. Otherwise, a new entry is created and labeled with a new UEs position set $\mathbf{s_{\mathrm{0}}}$. In Fig.~\ref{fig:bs_switching_rsm} it is assumed $L$ distinct UEs positions sets exist in the REM.

In addition to the UEs position set, every REM entry contains so-called, action values $Q(\mathbf{s},\action)$. These are metrics on how a particular active BSs set $\action$ is preferred, while a given UEs position set $\mathbf{s}$ is encountered. The methods of calculation and utilization of $Q(\mathbf{s},\action)$ are described in Sec.~\ref{sec:rl}.   
Finally, 
variable $N(\mathbf{s},\action)$ is stored for a given UEs location set $\mathbf{s}$ and a given active BSs set $\action$, presenting the number of times action $\action$ was taken in state $\mathbf{s}$.


\section{Reinforcement Learning} \label{sec:rl}
\begin{figure}[h]
\centering\includegraphics[scale=0.37]{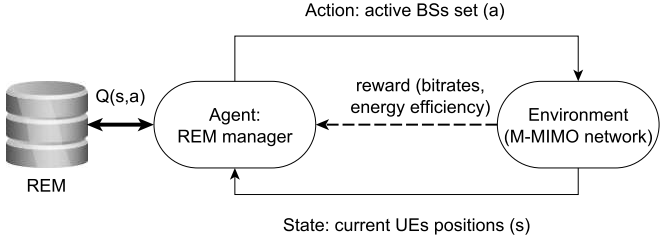}
\caption{Diagram of REM learning with the use of RL.}
\label{fig:reinforcement_learning}
\end{figure}

An efficient way to utilize REM data is to implement some RL algorithms. In RL the so-called agent learns through interaction with the so-called environment by taking appropriate actions and receiving the so-called reward~\cite{Whiteson2010}. The RL scheme utilized for increasing the energy efficiency of the considered M-MIMO HetNet by BSs switching is depicted in Fig.~\ref{fig:reinforcement_learning}. 

The consecutive elements of the RL framework will now be described in detail:
\begin{itemize}
    \item \textbf{State $\mathbf{s} \in \mathcal{S}$} is defined as the set of UEs coordinates rounded to fit the square grid of size $g$. States are recognized using the HD metric defined by~\eqref{eq:hd}. $\mathcal{S}$ is a set of all historical UEs positions stored in REM. 
    \item  \textbf{Action} $\action \in \mathcal{A}$ is a binary vector of $N_{\mathrm{BS}}$ elements. Each vector element represents one BS, and its value 1 or 0 represents its activity or sleep mode, respectively. All possible actions are denoted by set $\mathcal{A}$. As MBS will not be considered for switching off, i.e., the element of $\action$ representing MBS always equals 1, the size of action space $N_{\mathrm{a}}$ grows exponentially with the number of PBSs in the HetNet: $N_\mathrm{a} = 2^{N_{\mathrm{BS}}-1}$. However, it will be shown that $N_{\mathrm{a}}$ can be reduced using simple analytical estimations. 
    \item \textbf{Reward} $r(\mathbf{s}, \action)$ depends on EE and the number of served UEs. Commonly, the EE metric is computed by dividing the average UE bitrate (over all considered UEs in the network) by the average system power consumption \cite{tombaz2014}. In order to increase fairness, e.g., bitrate for cell-edge users, it is proposed to use median downlink user bitrate instead:
    \begin{equation}
     \label{eq:ee}
        EE(\mathbf{s}, \action) = \frac{c_{50}(\mathbf{s},\action)} {P_{\mathrm{avg}}(\mathbf{s},\action)},
    \end{equation}
        where $c_{50}(\mathbf{s},\action)$ is the median UE bitrate measured in state $\mathbf{s}$ while exploiting action $\action$, and  $P_{\mathrm{avg}}(\mathbf{s},\action)$ is the average power consumed by all BSs computed with \eqref{eq:p_tot}. In addition, reward $r(\mathbf{s},\action)$ depends on the number of currently served UEs. 
         On the basis of the channel estimates it can be deduced how many UEs can be served if all BSs were active $N_{\mathrm{UE}}(\mathbf{s},\mathbf{1})$ 
        , and compare it against $N_{\mathrm{UE}}(\mathbf{s},\action)$, being the number of UEs served in state $\mathbf{s}$ while taking action $\action$. Reward $r(\mathbf{s},\action)$ is thus given by:
    \begin{equation} \label{eq:reward}
    r(\mathbf{s},\action) = \begin{cases}
        EE(\mathbf{s},\action) & \text{ if } N_{\mathrm{UE}}(\mathbf{s},\action) \geq N_{\mathrm{UE}}(\mathbf{s}, \mathbf{1}) \\
        0 & otherwise \\
    \end{cases}.
\end{equation}
    This constraint causes that the active BSs set with the number of served UEs smaller than in a fully active network obtains a reward equal to 0. 
\item \textbf{Environment} is the considered M-MIMO HetNet. The environment respond with reward $r(\mathbf{s},\action)$ (as a result of transmission after previous action) and a new state $\mathbf{s}$ (as a result of UEs movement). An important observation is that modifications of the environment state, i.e., UEs position sets, are not influenced by the chosen actions, i.e., active BSs sets. This fact will be later used to simplify the RL problem.

\item \textbf{Agent} is the REM manager. Its goal is to maximize the overall future reward returned by the environment, by learning the environment through taking actions $\action$. We will use the so-called Q-learning algorithm, because it ensures convergence to the optimal solution, while requiring only simple updates of action values $Q(\mathbf{s},\action)$ stored in REM~\cite{sutton2018rl}. 

In Q-learning, every action $\action$ in state $\mathbf{s}$ can be characterized by an action-value $Q(\mathbf{s},\action)$ being the \emph{profit} of taking this action. 
Action values are updated according to the following rule:
\begin{equation} \label{eq:q_learning}
    Q(\mathbf{s},\action) \leftarrow Q(\mathbf{s},\action) + \alpha \cdot [r(\mathbf{s},\action) + \xi \cdot \max_{\action' \in \mathcal{A}} Q(\mathbf{s}',\action') - Q(\mathbf{s},\action)],
\end{equation}
where $\alpha$ is a step-size parameter, $\xi \in <0,1>$ is the discount-factor determining how much the action-value from future state $Q(\mathbf{s}',\action')$ should affect the current action value $Q(\mathbf{s},\action)$. As it was already mentioned, in the considered system all states are independent from the actions, i.e., UEs motion is not affected by the active BSs set. As a result, we can represent our problem as a set of independent optimization tasks in each state. However, solving them still requires RL to interact with the environment. Such a simplified RL problem is known in the literature as \emph{Associative Search}, or \emph{Contextual Bandit}~\cite{sutton2018rl}. We decided to describe it equivalently as "myopic" Q-learning, with the discount-factor set to zero: $\xi = 0$. Thus, the update of the action values is simply a $1-\alpha$ filter: 
\begin{equation} \label{eq:update_q}
    Q(\mathbf{s},\action) \leftarrow (1-\alpha) \cdot Q(\mathbf{s},\action) + \alpha \cdot r(\mathbf{s},\action).
\end{equation}
Therefore, in this particular case, $Q(\mathbf{s},\action)$ can be found as the forecasted reward for taking action $\action$ in state $\mathbf{s}$. 
\end{itemize}

\subsection{Exploration Strategies}
One of the RL challenges is to properly balance the time spent by the algorithm on exploiting the current knowledge, and exploring new actions. There are several exploration-exploitation strategies, widely used in order to manage action selection on the basis of action values $Q(\mathbf{s},\action)$.

\subsubsection{$\epsilon$-greedy} \label{subsubsec:e_greedy}

The simplest approach is the so-called $\epsilon$-greedy \cite{tijsma2016}. The algorithm is driven by the $\epsilon \in \langle 0;1 \rangle$ parameter, being the probability of exploration. Encountering state $\mathbf{s}$, action $\action_t = \arg \max_{\action \in \mathcal{A}}Q(\mathbf{s},\action)$, i.e., a greedy action, is chosen with probability of $1-\epsilon$. With the probability of $\epsilon$, a random action $\action_t \in \mathcal{A}$ is chosen. It seems reasonable to first focus more on the exploration, while later on the exploitation. Therefore, it is proposed that $\epsilon$ be calculated using:
\begin{equation}
    \label{eq:epsilon_greedy}
    \epsilon = \frac{1}{\sqrt[\beta]{\sum_{\action \in \mathcal{A}}N(\mathbf{s},\action)}},
\end{equation}
where $N(\mathbf{s},\action)$ was defined in Sec. \ref{subsec:rem_data_structure} and $\beta$ is the weighting factor allowing for balance between exploitation and exploration. 
\subsubsection{Upper Confidence Bound} \label{subsubsec:ucb}
The Upper Confidence Bound (UCB) is a more advanced exploration strategy~\cite{sutton2018rl}. Its advantage is that it does not utilize randomness. Action $\mathbf{a_t}$ is selected on the basis of action value $Q(\mathbf{s},\action)$, the number of visits in state $\mathbf{s}$, i.e., $\sum_{\mathbf{k} \in \mathcal{A}}N(\mathbf{s},\mathbf{k})$, and the number of times that a particular action was selected $N(\mathbf{s},\action)$:
\begin{equation} 
    \mathbf{a_t} = \arg \max_{\action \in \mathcal{A}} \left\{Q(\mathbf{s},\action) + c \cdot \sqrt{\frac{\ln{\sum_{\mathbf{k} \in \mathcal{A}}N(\mathbf{s},\mathbf{k})}}{N(\mathbf{s},\action)}} \right\},
    \label{eq:ucb}
\end{equation} 
where $c$ is an arbitrary constant weighting the exploration-exploitation strategies. If $c$ is high, then the algorithm would often choose rarely explored actions. On the other hand, the lower the $c$, the more greedy the algorithm becomes. 

Both UCB and $\epsilon$-greedy algorithms can benefit from the so-called \emph{optimistic initialization}, i.e., setting the initial values of $Q(\mathbf{s},\action)$ to a high one. The algorithm is then forced to test every action at least once. If the set of possible actions is large, it prolongs the convergence of the algorithm until all actions are tested at least once. 

\subsubsection{Gradient Bandit} \label{subsubsec:gb}
The Gradient Bandit (GB), in contrast to the $\epsilon$-greedy and UCB algorithms, relies not on the direct reward values but on relations between them, by obtaining  a probability distribution of taking each action \cite{sutton2018rl}. The probability of taking action $\action$ follows the so-called soft-max distribution:
\begin{equation}
    P\{\action_t =\action\} = \frac{e^{Q(\mathbf{s},\action)}}{\sum_{\mathbf{k} \in \mathcal{A}} e^{Q(\mathbf{s},\mathbf{k})}} = \pi (\action).
\end{equation}
The formula for updating the action value is slightly different from \eqref{eq:update_q}:
\begin{equation}
\begin{cases}
    Q(\mathbf{s},\action_t) \leftarrow Q(\mathbf{s},\action_t) + \alpha_{\mathrm{gb}} \{r(\mathbf{s},\action_t) - \hat{r(\mathbf{s})} \}  \{ 1 - \pi(\action_t)\}, & \text{and}  \\
        Q(\mathbf{s},\action) \leftarrow Q(\mathbf{s},\action) - \alpha_{\mathrm{gb}} \{r(\mathbf{s},\action) - \hat{r(\mathbf{s})} \} \cdot   \pi(\action), & \text{for }  \action \neq \action_t 
\end{cases},    
\end{equation}
where $\alpha_{\mathrm{gb}}$ is the step-size parameter, and 
$\hat{r(\mathbf{s})}$ is the average reward received in state $\mathbf{s}$ in the previous visits. The main idea is that if the chosen action provides a higher reward, then the probability of its exploitation increases.
The \emph{optimistic initialization} in the case of GB doesn't work, because the action values are relative to each other in this case.  

\subsection{Proposed Action Space Reduction (ASR) Method} \label{subsec:asr}

Frequent switching BSs on/off is not encouraged by wireless network operators and users. Operators are afraid ofdestabilizing or damaging the hardware installed on BSs. Users don't want to get poor Quality of Experience (QoE). Therefore, it is reasonable to reduce the RL convergence time, e.g., by \emph{a priori} removal of actions that cannot be optimal in a given state.
It is proposed that the actions which result in zero-reward be excluded. Suppose that there is a UE that receives a strong enough signal (to be served) from only a single BS. This BS should be explicitly prevented from being switched off to preserve the required QoS, i.e., not limiting the number of connected UEs. 

Due to the channel hardening property of the M-MIMO, the average channel gain is relatively stable despite the slight variations of every single radio channel \cite{massivemimobook}. Thus on the basis of channel estimates in  state $\mathbf{s}$ we can \emph{a priori} estimate the number of UEs having RSS above the threshold required for communication with minimum quality with any of the active BSs set (action) $N_{\mathrm{ue}}(\mathbf{s,\action})$:
\begin{equation}
    N_{\mathrm{ue}}(\mathbf{s},\action) = \sum_{s_i \in \mathbf{s}}f(s_i,\action),
\end{equation}
where $s_i$ denotes the localization of UE $i$, and $f(s_i,\action)$ determines if UE $i$ can be served by the network while considering action $\action$:
\begin{equation}
    f(s_i,\action) = \begin{cases} 1 & \text{if UE $i$ can be served}, \\  0 & \text{otherwise}, 
    \end{cases}
\end{equation}

 Using the above equations, the number of served users $N_{\mathrm{ue}}(\mathbf{s},\action)$ related to each action $\action \in \mathcal{A}$ can be estimated before making the decision about BSs switching.
Every action that results in a reduction of the number of served users in comparison to the maximal number $\max_{\action \in \mathcal{A}}\{N_{\mathrm{ue}}(\mathbf{s},\action)\}$ (over all actions) is excluded from the action space (never explored), i.e., BSs crucial for providing the minimal QoS are explicitly prevented from being switched off. As a result of ASR a smaller action space $\mathcal{\Grave{A}}$ is obtained  and used in the algorithms instead of $\mathcal{A}$, i.e., $\action \in \Grave{\mathcal{A}}$. 

\subsection{REM-based Exploration Algorithm (REM-EA) } \label{subsec:rem_ea}
Although we concluded that the transitions between states are not influenced by  actions, we can expect that a similar optimal active BSs set, i.e., action $\action$, should be chosen in the similar states, i.e., similar UEs position sets. On the basis of promising initial results and observations of the UCB algorithm, our second proposal is to modify it  
so as to exploit the knowledge from all REM entries, i.e., states, but properly weighted. Most importantly, this is executed with reduced action space, i.e., after using ASR. Let us define $\mathbf{s}_0$ as the currently visited state and $\mathbf{s}_l$ as another state related to REM entry $l$. The REM-EA is then given as: 
\begin{equation} 
    \action_t = arg \max_{\action \in \Grave{\mathcal{A}}} \left\{\hat{Q}(\action) + c \cdot \sqrt{\frac{\ln{\sum_{\mathbf{k} \in \Grave{\mathcal{A}}}\hat{N}(\mathbf{k})}}{\hat{N}(\action)}} \right\},
    \label{eq:rem_ea}
\end{equation}
where:
\begin{equation}
    \hat{Q}(\action) = \frac{Q(\mathbf{s}_0,\action) + \sum_{\mathbf{s} \in \mathcal{S}\setminus \mathbf{s}_0, Q(\mathbf{s},\action) \neq 0} \frac{Q(\mathbf{s},\action)}{d_h(\mathbf{s}_0,\mathbf{s})^\gamma} }{1 + \sum_{\mathbf{s} \in \mathcal{S}\setminus \mathbf{s}_0, Q(\mathbf{s},\action) \neq 0} \frac{1}{d_h(\mathbf{s}_0,\mathbf{s})^\gamma}},
\end{equation}
\begin{equation}
    \hat{N}(\action) = \frac{N(\mathbf{s}_0,\action) + \sum_{\mathbf{s} \in \mathcal{S}\setminus \mathbf{s}_0} \frac{N(\mathbf{s},\action)}{d_h(\mathbf{s}_0,\mathbf{s})^\gamma} }{1 + \sum_{\mathbf{s} \in \mathcal{S}\setminus \mathbf{s}_0} \frac{1}{d_h(\mathbf{s}_0,\mathbf{s})^\gamma}},
\end{equation}
and $d_h(\mathbf{s}_0,\mathbf{s})$ is the HD between the UE position set $\mathbf{s}_0$ and $\mathbf{s}$ \eqref{eq:hd}, and $\gamma$ is an arbitrary constant. The larger the $\gamma$, the smaller the impact on the current action selection more distanced REM entries have.  
It will be shown in the computer simulation that the proposed reasoning based on spatial (in states space) correlation, i.e., REM-EA, can provide further improvement for RL convergence, due to the efficient utilization of REM data. 

\section{Simulation Results} \label{sec:simulation}

The algorithms described in the previous sections are evaluated in terms of computer simulations using a realistic, 3D-ray-tracing radio channel model. We consider a system level simulation of a single, two-tier M-MIMO HetNet composed of an MBS and a set of PBSs. Each resembles a 5G system by using proper configuration, e.g., OFDM with numerology, slots and subframes, channel estimation using uplink pilots, Modulation And Coding Schemes and effective SINR mapping. The M-MIMO uses a Regularized Zero Forcing precoder \cite{massivemimobook}. The downlink traffic is used for the optimization of HetNet EE and presented using various statistics in later sections.

\subsection{Simulation Environment}
The implemented simulator consists of several functional blocks that will be briefly described in the next subsections.
\subsubsection{Base Station Deployment}
\begin{figure}[h]
\centering\includegraphics[width=9.1cm]{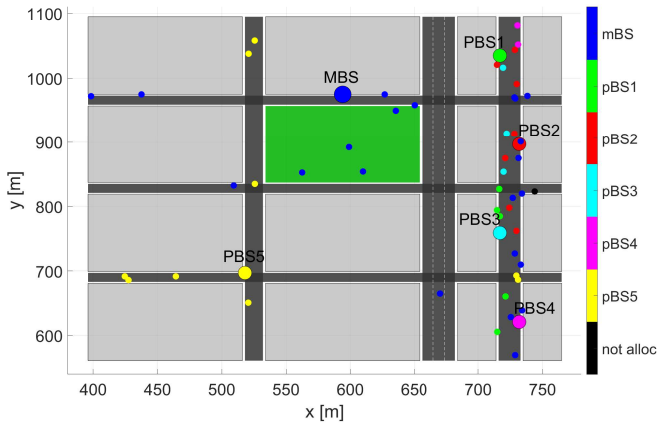}
\caption{Deployment of the BSs (big dots) in the considered M-MIMO HetNet, and UEs initial positions (small dots - color-coded is association of a UE to a BS).}
\label{fig:ue_deployment}
\end{figure}

The HetNet is deployed in an urban scenario following the \emph{Madrid Grid Model}, which exploits various challenging regions: a park, narrow street canyons, and a promenade~\cite{metis2020}. There are in total 6 BSs in the considered system, i.e., 1 MBS and 5 PBSs. The BSs are deployed on the basis of initial coverage simulations as it is shown in  Fig.~\ref{fig:ue_deployment}. The MBS is placed in the central part of the map near the park, and provides general coverage. PBSs 1 to 4 are located along the promenade, aiming at the improvement of network capacity in the most crowded area. PBS 5 is placed in the street canyon to extend the network coverage in the place where MBS coverage is insufficient.
Both the MBS and PBSs are equipped with rectangular antenna arrays. The MBS exploits a single 128-element panel arranged in a $16 \times 8$ rectangle, i.e., having 16 columns and 8 rows. 

Each PBS is equipped with two $2\times 8$ (2 columns, 8 rows) rectangular panels pointing in opposite directions. The MBS antenna array is deployed 2.5 m above the rooftop at the total height of 45 m. The PBS antenna arrays are claimed to be installed on street lamps 5 m above the ground.


\subsubsection{UE to BS Association}

In the considered network, the UE-to-BS association is centralized using the Dynamic Point Selection (DPS) concept. It is performed every 10~ms. In order to reduce the system complexity, and focus only on BSs switching, a simple UE-to-BS association based on the Received Signal Strength (RSS) is used. RSS from a given UE is summarized over all antennas available at a given BS. The UEs are not associated with any BS, if their highest (over BSs) average (over antennas) RSS is below the threshold $\mathrm{P_{th}}=-120 dBm$. This threshold is based on minimal MCS requirements according to the parameters specified in Sec. \ref{sec:sched_prec}.

There are 50 UEs generated, and their initial positions are depicted in Fig. \ref{fig:ue_deployment}. They are generated so that the most crowded area is the promenade in the right part of the considered map. However, some UEs are located in the park and in the street canyons as well. The UEs move with the speed of $1.5 \frac{m}{s}$.


\subsubsection{Traffic Model}

In this paper, the data stream to each UE follows the \emph{full-buffer} model, i.e., even a single UE associated to a BS utilizes all of its time-frequency resources. As a result, the traffic load in the considered M-MIMO HetNet is always maximal, creating the most challenging scenario for EE improvement. 

\subsubsection{Scheduler, Precoder}
\label{sec:sched_prec}
For users associated to a given BS, the scheduling and precoding is executed every timeslot. The scheduler is based on the design proposed in \cite{Goldsmith_scheduling_MIMO_2006}. For each Resource Block (RB), independent initial allocation is performed. Associated UEs are sorted according to their proportionality fairness metric, i.e., the ratio of potential rate and past rate. The considered M-MIMO utilizes Regularized Zero Forcing Precoder that allows many UE layers to be allocated in parallel to the same time-frequency resources. However, the scheduler rejects a new UE whose channel is too correlated with already allocated UEs (based on the channel correlation coefficient) and stops adding layers when the total estimated bitrate decreases. The SINR values estimated for each RB allocated to a given UE are used for MCS selection. This utilizes Exponential Effective SINR mapping and 15 MCSs with proper coefficients defined in \cite{Bossy_MCS_LTE, Hanzaz_MCS_LTE}.

\subsubsection{Channel Generation} \label{subsubsec:channel_generation}

\begin{figure}[h]
\centering\includegraphics[scale=0.2]{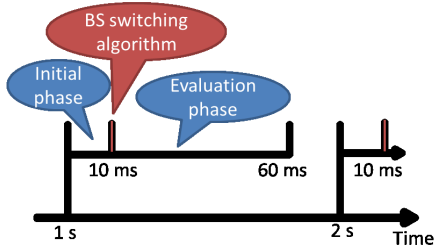}
\caption{Channel generation scheme.}
\label{fig:channel}
\end{figure}

The 3D ray-tracing model is claimed to be the most accurate and realistic for the evaluation of the M-MIMO network \cite{shang2016}. The cost of the utilization of a 3D ray-tracing model is the long time of channel generation, large amount of required storage, and long simulation time itself. To overcome these limitations, and allow the simulation of relatively long UEs paths, we propose to generate the radio channel coefficients in several short batches, as it is depicted in Fig.~\ref{fig:channel}. The channel is generated and system-level simulations are carried out only over each batch period.  

There are 15 batches generated, each lasting 60 ms. There is a 1 s time gap between the consecutive batches. It is to allow the UEs to change their positions significantly, which can result in a different optimal active BSs set. First, 10 ms of each batch is for initiating the simulator, e.g., to obtain initial UE rates or interference statistics, then the BS switching algorithm is launched and the obtained active BSs set (action) is evaluated during the last 50 ms. The statistics are obtained based on the last 50 ms of each batch. To provide some changes and randomness in the radio environment, there are 10 different channels generated with the same UEs paths but different locations of scatterers in the environment, e.g., persons. In addition, channel variation is boosted by Gaussian channel estimation error generated independently for each simulation iteration, following the model described in~\cite{serra2012chester}.  

\subsection{Simulation Setup}

The main parameters of the simulation environment are presented in Tab. \ref{tab:simulation_setup}. Every simulation is performed using the same setup by default. Any slight changes will be explicitly described in the text.
\begin{table}[h]
\centering
\begin{tabular}{l l}
\hline
\textbf{Parameter} & \textbf{Value} \\
\hline
Simulation Time & ~15 s \\ Time Slot Duration & 0.5 ms \\ Number of UEs & 50 \\
Channel Realizations & 10 \\ UEs Speed & $1.5 \frac{m}{s}$ \\
Number of PBSs & 5 \\ Central Frequency & $3.55$ GHz \\
Bandwidth & 300 MHz \\ Subcarrier Spacing & 30 kHz \\
Number of MBS Antennas & $128$ ($16\times 8$) \\
Number of PBS Antennas & $32$ ($2\times 2 \times 8$) \\
RSS threshold $P_{\mathrm{th}}$ & -120 dBm \\
REM Grid Size $g$ & 3 m \\
\hline
\end{tabular}
\caption{Simulation system parameters}
\label{tab:simulation_setup}
\end{table}

The parameters related to the Power consumption model from Section \ref{subsec: power_model} are presented in Tab. \ref{tab:power_model}.
\begin{table}[h]
\centering
\begin{tabular}{l l}
\hline
\textbf{Parameter} & \textbf{Value} \\
\hline
MBS Transmitted Power $P_{\mathrm{tx,MBS}}$ & 46 dBm \\
PBS Transmitted Power $P_{\mathrm{tx,PBS}}$ & 30 dBm \\
Amplifier Efficiency $\eta$ & 0.5 \cite{massivemimobook}\\
Transceiver Chain Power $P_{\mathrm{\widehat{TC}}}$ & 0.4 W \cite{massivemimobook}\\
Local Oscillator Power $P_{\mathrm{LO}}$ & 0.2 W \cite{massivemimobook} \\
BS Fix Power $P_{\mathrm{fix}}$ & 10 W \cite{massivemimobook} \\
BS Stand-By Power $P_{\mathrm{off}}$ & 10 W \cite{frenger2011} \\

\hline
\end{tabular}
\caption{Power model parameters}
\label{tab:power_model}
\end{table}

\subsection{Benefits of the REM Solution}

The first simulation experiment is performed in order to show the maximum potential benefits of the utilization of REM, compared to the existing solution based on analytical heuristics. The existing analytical heuristic is the so-called switching on/off-based energy saving (SWES) algorithm \cite{oh2013swes}. The SWES algorithm has been adopted to the \emph{full-buffer} UE traffic model. Originally, SWES switches off BSs according to their traffic load. We modified SWES to switch BSs on/off on the basis of the median user bitrate obtained  with the Shannon formula. A BS can be switched off if the initial median user bitrate is degraded by no more than $5\%$.

To obtain the REM solution, we temporarily switched off the radio environment variations, i.e., channel estimation error, and exploited only single channel realization. Then, having an invariable radio environment, the REM was learned using an exhaustive search. 

\begin{figure}[h]
\centering\includegraphics[scale=0.27]{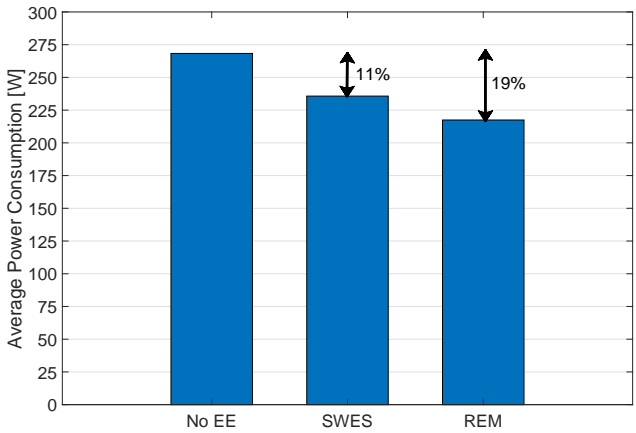}
\caption{Achievable energy savings, while there is no EE optimization (no ee), for SWES algorithm, and REM, respectively.}
\label{fig:rem_power}
\end{figure}

\begin{figure}[h]
\centering\includegraphics[scale=0.27]{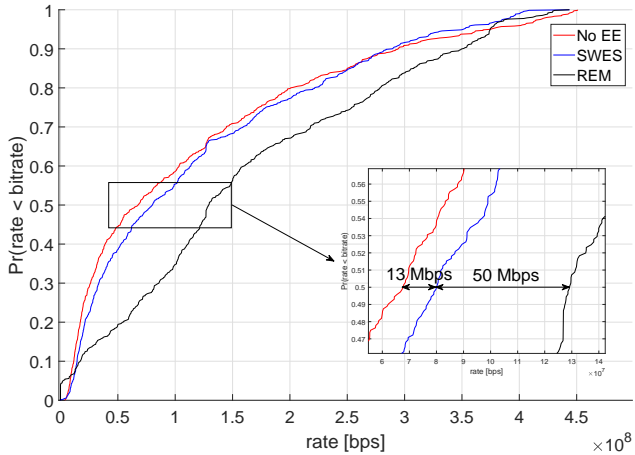}
\caption{Empirical Cumulative Distribution Function of UEs bitrates, while there is no EE optimization (no ee), for SWES algorithm and REM respectively.}
\label{fig:rem_cdf}
\end{figure}

 The resultant energy savings are depicted in Fig.~\ref{fig:rem_power}. It can be observed that the REM solution performs better than the SWES algorithm, increasing the overall power savings to 19\%. But not only the power saving are higher due to the utilization of REM solution. In Fig.~\ref{fig:rem_cdf} it is shown that the median user bitrate is about 50 Mbps higher when utilizing REM in relation to the SWES algorithm. At the same time, SWES provides a median user bitrate gain of about 13 Mbps over a scenario without EE optimization. The bitrate gains are caused by reduced interference related to switching off some of the BSs, which remains consistent with the previous research findings~\cite{ge2015}. The interference changes are hard to be analytically estimated, thus SWES switches off the sub-optimalset of BSs. Taking into the account the gains from energy savings and median user bitrate, the overall system EE is 70\% higher when utilizing REM in comparison to the SWES algorithm. Moreover, EE obtained with the REM solution is 136\% higher in relation to the scenario without EE optimization.

\subsection{State-of-the-Art Exploration Algorithms}

We have shown that REM can provide the network with s much better solution than SWES. The challenge, is however, to efficiently learn, i.e., minimize, the number of times a given action has to be tested in a given state $\mathbf{s}$ before stabilizing at one solution, hopefully, the best one. Moreover, the utilized RL has to be able to adapt to changing radio environment conditions. First, the state-of-the-art exploration strategies: $\epsilon$-greedy (Sec.~\ref{subsubsec:e_greedy}), UCB (Sec.~\ref{subsubsec:ucb}), and Gradient Bandit (Sec.~\ref{subsubsec:gb}) are studied to obtain the best parameters for fair comparison and application of the improvements proposed in Sec. \ref{subsec:asr} and Sec. \ref{subsec:rem_ea}.

\subsubsection{$\epsilon$-greedy}
\begin{figure}[h]
\centering\includegraphics[scale=0.27]{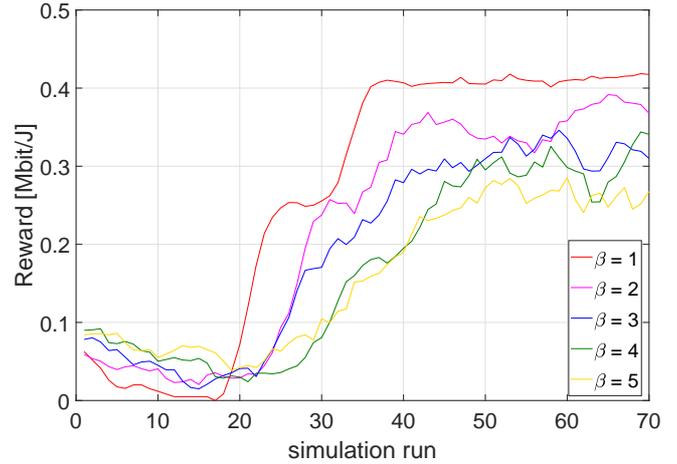}
\caption{Convergence of the $\epsilon$-greedy exploration method. There is a 5-tap moving average applied over simulation runs.}
\label{fig:e_greedy_convergence}
\end{figure} 

Fig.~\ref{fig:e_greedy_convergence} depicts a convergence of the $\epsilon$-greedy exploration method. It can be seen that the lower the $\beta$ parameter, the faster the convergence is. Low $\beta = 1$, being the optimal value, indicates that it is profitable for the algorithm to perform greedily. For a higher $\beta$ value, the algorithm spends too much time on the exploration of random actions, which are often related to a low reward.

\subsubsection{UCB}
\begin{figure}[h]
\centering\includegraphics[scale=0.27]{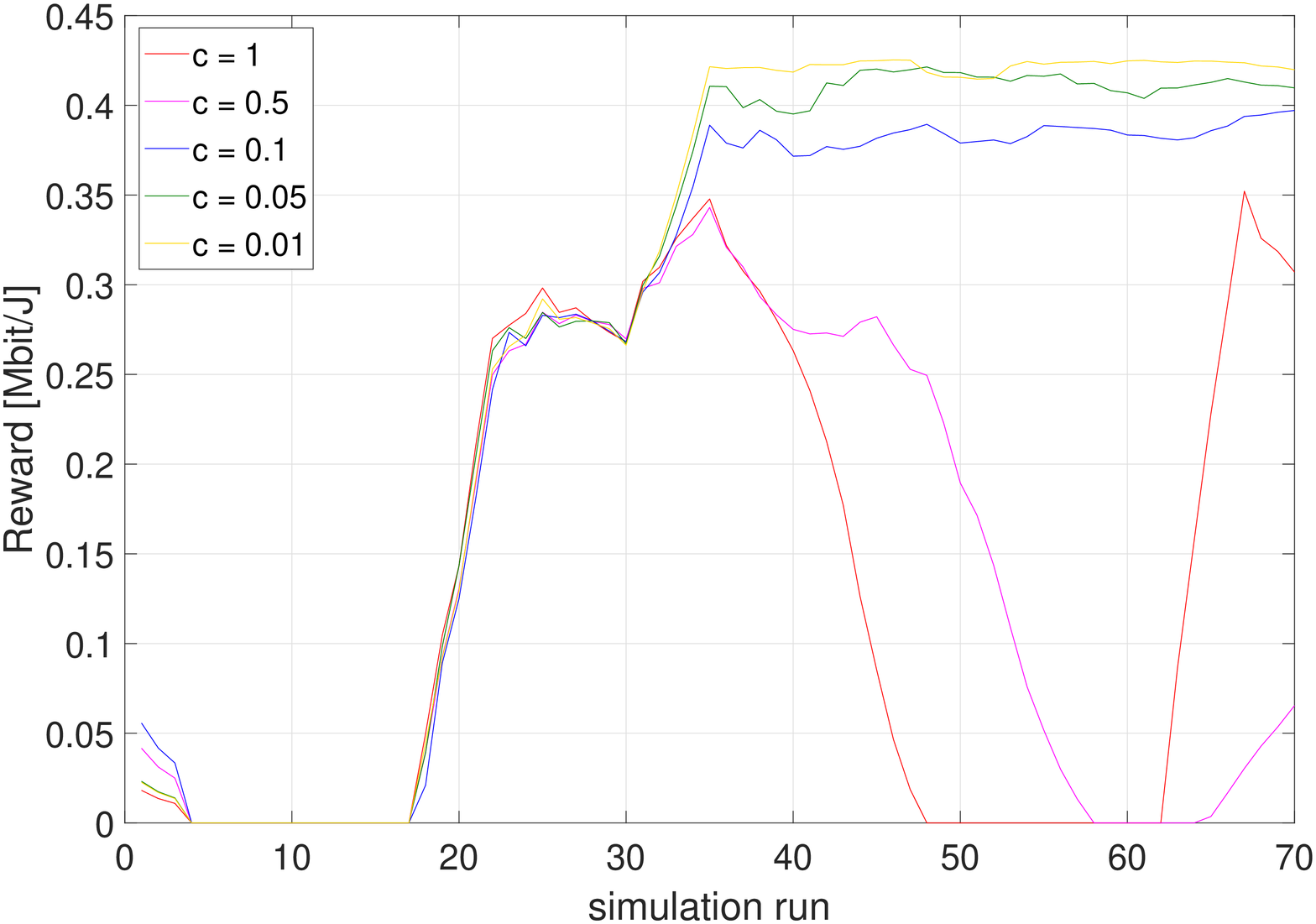}
\caption{Convergence of the UCB exploration method. There is a 5-tap moving average applied over simulation runs.}
\label{fig:ucb_convergence}
\end{figure}

A similar tendency can be observed in the convergence of the UCB algorithm depicted in Fig.~\ref{fig:ucb_convergence}. The $c$ parameter in \eqref{eq:ucb} provides a balance between executing a greedy action and exploring less known actions. For high $c$, the actions providing a low reward are chosen too often. We can see that in the case of UCB also the algorithm performs better when it is focused mainly on the greedy action exploitation. The chosen value of $c$ equals $0.01$. 

\subsubsection{Gradient Bandit}

\begin{figure}[h]
\centering\includegraphics[scale=0.27]{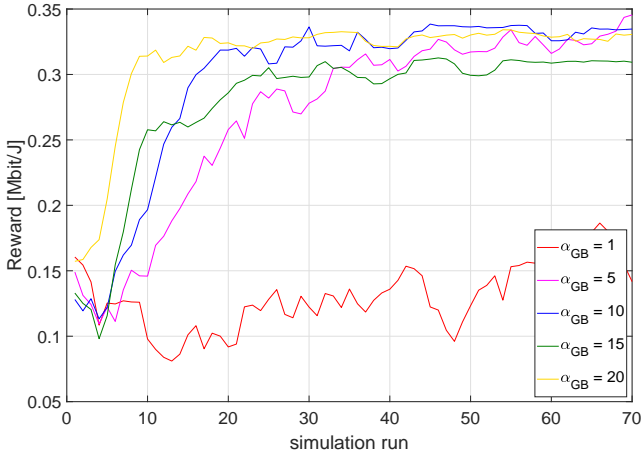}
\caption{Convergence of the Gradient Bandit exploration method. There is a 5-tap moving average applied over simulation runs.}
\label{fig:gb_convergence}
\end{figure}

The Gradient Bandit algorithm performs on the basis of the probability distribution of taking particular actions. Action probability increases when the reward is higher than the average reward. As we can see in Fig.~\ref{fig:gb_convergence}, the algorithm has a tendency to run into the sub-optimal solution. However, the fastest convergence could be observed for $\alpha_{\mathrm{GB}} = 20$. 

\subsubsection{Comparison and Application of ASR}
\begin{figure}[h]
\centering\includegraphics[scale=0.27]{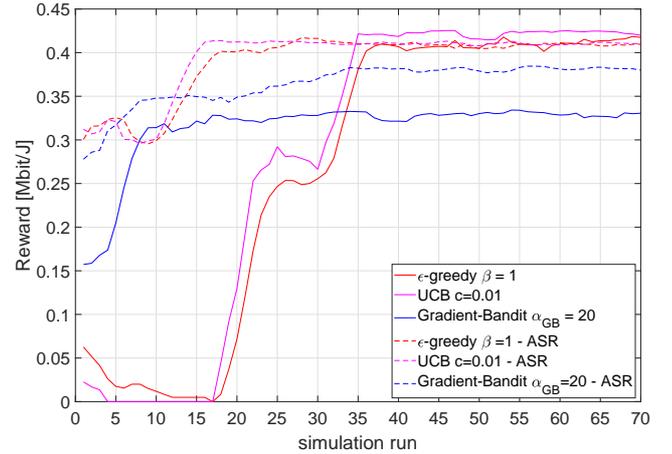}
\caption{Comparison of the state-of-the-art exploration methods (solid lines), and improvement using ASR (broken lines). There is a 5-tap moving average applied over simulation runs.}
\label{fig:comparison_convergence}
\end{figure}

Additionally, all of the three exploration methods are evaluated in terms of the mean of the reward obtained on the basis of the last 30 simulation runs. The results are presented in Tab.~\ref{tab:comparison_mean}. It can be observed that Gradient Bandit does not converge to the solution providing a maximal reward visible in the $\epsilon$-greedy and UCB methods. The comparison of the algorithms' convergence for the best previously obtained parameters is depicted in Fig.~\ref{fig:comparison_convergence}. The fastest convergence can be observed for the Gradient Bandit algorithm, however it stabilizes at a suboptimal solution. The $\epsilon$-greedy and UCB perform similarly, with the slight advantage on the side of UCB. These converge after about 35 simulation runs. The  broken line in Fig.~\ref{fig:comparison_convergence} depicts the result of the ASR (Sec.~\ref{subsec:asr}) applied to the considered RL algorithms. It can be seen that in the case of UCB the convergence can be reduced from about 35 to about 15 simulation runs, i.e., about 2.3 times. This improved is obtained thanks to reduction of actions space. In the next section it will be shown that the convergence speed can be further increased using knowledge from another REM entries (different states).     
\begin{table}[h]
\centering
\begin{tabular}{l l | l l | l l }
\hline
\multicolumn{2}{c|}{$\epsilon$-greedy} &
\multicolumn{2}{c|}{UCB} &
\multicolumn{2}{c}{Gradient Bandit}\\
\textbf{$\beta$} & EE $(\frac{Mbit}{J})$ &c & EE $(\frac{Mbit}{J})$ & $\alpha_{\mathrm{GB}}$ & EE $(\frac{Mbit}{J})$ \\
\hline
1 & 0.41 & 1 & 0.14  & 1 & 0.14 \\
2 & 0.35  & 0.5 & 0.11 & 5 & 0.32 \\
3 & 0.31  & 0.1 & 0.38 & 10 & 0.33  \\
4 & 0.28  & 0.05 & 0.41 & 15 & 0.30  \\
5 & 0.26  & 0.01 & 0.41 & 20 & 0.33 \\
\hline
\end{tabular}
\caption{Mean reward obtained in the last 30 simulation runs, for $\epsilon$-greedy, UCB, and Gradient Bandit exploration algorithms}
\label{tab:comparison_mean}
\end{table}

\subsection{Performance of REM-EA}

\begin{figure}[h]
\centering\includegraphics[scale=0.27]{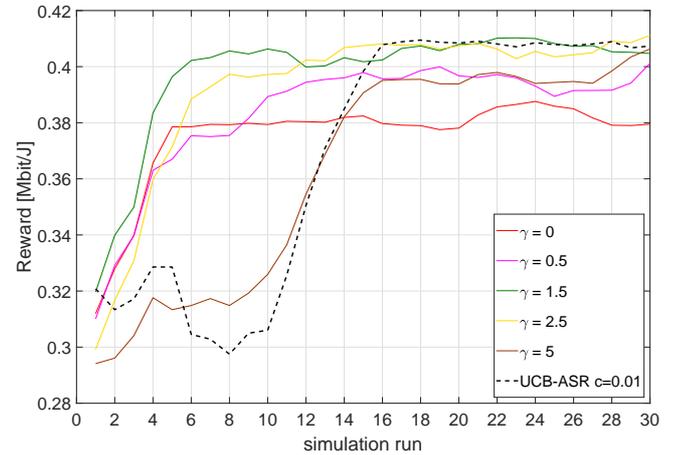}
\caption{REM-EA convergence over different $\gamma$ values. There is a 5-tap moving average applied over simulation runs.}
\label{fig:rem_ea_convergence}
\end{figure}

Finally, the REM-EA from Sec.~\ref{subsec:rem_ea} is evaluated, to show that it is beneficial for the RL algorithms to utilize information stored in REM for various states (UEs positions) to build a knowledge about the current state. REM-EA is based on the UCB as it has been proved in the previous sections to provide the best results. The same constant $c=0.01$ is used for REM-EA. However, REM-EA has in addition a parameter $\gamma$ weighting the impact of distanced REM information  (in state sense) on the action being chosen in the current state. The results of REM-EA for various $\gamma$ values are presented in Fig.~\ref{fig:rem_ea_convergence}. The results are compared against the UCB algorithm with ASR applied. 
For high $\gamma$ values, REM-EA does not exploit the knowledge from similar REM entries and tends to perform as the pure UCB algorithm. However, for low $\gamma$ values the impact of neighbouring states is too high, which results in reaching only a sub-optimal solution. The optimal $\gamma$ value providing the proper weighting of REM data seems to be 1.5. The application of the REM-EA can speed up the convergence to about 6 simulation runs, i.e., in the considered network the proposed algorithm requires 6 passes of UEs, each lasting 15 s, to obtain enough knowledge to find an energy-efficient BSs activity pattern. It is a convergence time reduction by about 60\%  in relation to the UCB with ASR, and about 83\% in relation to the pure UCB, while remaining the same optimal solution. Obviously, this is an example of starting from an empty REM, i.e., worst case scenario. If there is some knowledge gathered in the REM the EE solution is claimed to  be obtained much faster than in 6 simulation runs. However, REM already containing a huge number of entries would be required to prove this.

\begin{figure}[h]
\centering\includegraphics[scale=0.27]{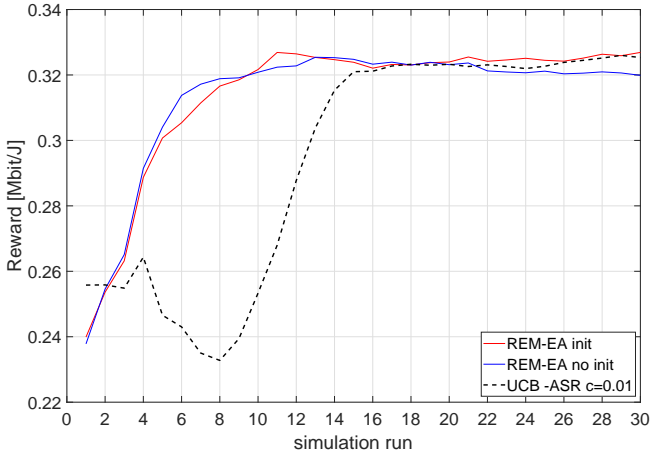}
\caption{Convergence of REM-EA while there are entries in REM (REM-EA init), REM-EA starting form empty REM (REAM-EA no init) and UCB, under a scenario with 45 UEs. For $c=0.01$, and $\gamma=1.5$. There is a 5-tap moving average applied over simulation runs.}
\label{fig:rem_ea_45ue}
\end{figure}

To evaluate how REM-EA performs while REM already has some entries and unknown state appears, a single channel realization exploiting 45 instead of 50 UEs is considered. Each of UEs locations, together with the radio channel coefficients are independently randomly generated, with the same spatial density. The REM-EA algorithm has been run in three cases: with initial knowledge from the previous simulations, without knowledge, and compared against the UCB-ASR method. The results are depicted in Fig.~\ref{fig:rem_ea_45ue}. The most crucial is the result for the REM-EA init approach. Its performance is very similar to running REM-EA without previous knowledge. It seems that previous knowledge saved in REM is too distanced (in terms of HD) to boost convergence. Most importantly, the previous knowledge degrades neither the convergence speed nor the maximal reward. Still, the convergence time is over 50\% faster in comparison to the UCB algorithm. REM-EA can adopt to the new environment conditions without performance degradation. 

\subsection{Square Grid Size Analysis}

\begin{figure}[h]
\centering\includegraphics[scale=0.27]{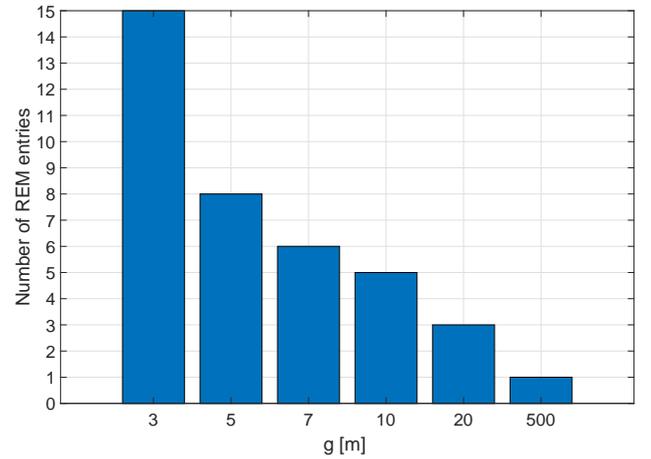}
\caption{The number of REM entries over various square grid size $g$.}
\label{fig:rem_entries}
\end{figure}

We propose in Sec.~\ref{sec:rem} to use a square grid of size $g$ in our REM, to store the REM entries labeled with UEs position set $\mathbf{s}$. 
While previously relatively high accuracy was used, i.e., $g=3 m$, this can result in a significant REM size. Now, coarser grid sizes $g$, resulting in the reduction of the number of REM entries, are evaluated. 
\begin{figure}[h]
\centering\includegraphics[scale=0.27]{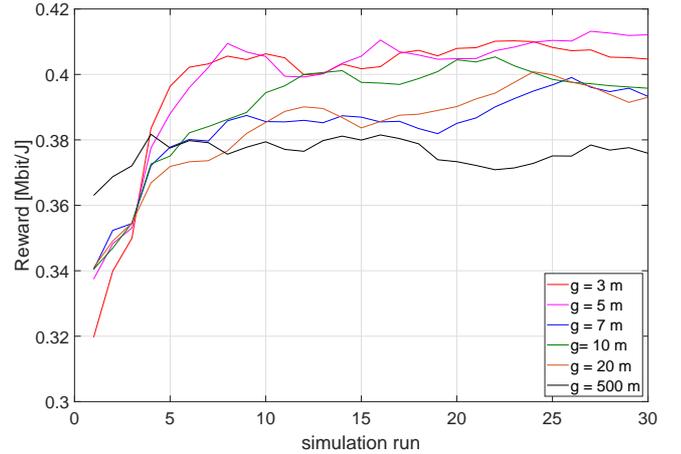}
\caption{REM-EA convergence over various square grid size $g$ for $\gamma=1$, and $c=0.01$. There is a 5-tap moving average applied over simulation runs. 
}
\label{fig:rem_ea_grid}
\end{figure} 
It can be seen in Fig.~\ref{fig:rem_ea_grid} that lowering the grid resolution does not increase the RL convergence. However, larger grid size (above $g=5 m$) results in reduced maximal reward value. This is caused by the smaller flexibility of BSs switching, i.e., for a coarser grid significantly different UEs propagation conditions require a single BSs activity pattern to be chosen.
Fig.~\ref{fig:rem_entries} presents the number of REM entries related to each grid resolution. The conclusion is that there is a trade-off between the number of REM states and REAM-EA performance. There can be found a grid resolution, e.g., $g=5$ m, which can provide both a high reward and minimized REM size, i.e., a reduction by almost 50\% in comparison to a system with $g=3 m$.  
\section{Conclusions} \label{sec:conclusions}

This paper proposes location-dependent data for providing EE gains in the M-MIMO HetNet. The underutilized BSs are switched off on the basis of the data stored in REM. First, the design of the REM and its deployment in the HetNet is proposed. Secondly, RL is proposed to manage the data acquisition and utilization process, using state-of-the-art exploration algorithms, i.e., $\epsilon$-greedy, UCB, and Gradient Bandit. Finally, two algorithms are proposed to speed up the RL convergence time. In ASR the number of served UEs is estimated on the basis of an analytical formula that allows the reduction of the action space. The second proposed algorithm is REM-EA. The data from the whole REM are utilized to obtain a proper active BSs set in the current state. The algorithms were extensively evaluated using an advanced M-MIMO HetNet simulator, and a realistic 3D-ray-tracing radio channel model. It has been proven that REM-based solution performs better then the state-of-the-art SWES algorithm. Further studies have shown that the state-of-the-art RL exploration algorithms' convergence benefits first, from the analytical estimations, i.e., ASR, and secondly, from the utilization of the similarities between the states stored in REM, i.e., REM-EA. 

In the future, the current study can be extended to include higher-mobility UEs or possibly vehicles. Although in the article the problem was formulated as \emph{Associative Search}, where the reward, i.e., EE, can be maximized in each state independently, it may be worth considering the dependency between the consecutive states in the future. Such a dependency can occur while taking into account the fact that frequent BS on/off switching increases the hardware expenditure~\cite{chiaraviglio2015}. One interesting further study is to incorporate both the costs of energy and those related to potential hardware replacement in the reward function. The aim of the agent would then be to minimize the operator cost, and balance the savings from EE BS switching, and expenditures related to hardware performance degradation. In that case, it would also be necessary to study the impact of the learning factor on the long-term Q-learning performance. Finally, studies on the reduction of REM size without performance degradation are important from the point of view of algorithm implementation in a real network. This can be achieved, e.g., by utilizing deep neural networks, but requires long simulation runs and varying UEs paths to first populate the REM. 

\section*{Acknowledgement} 
The presented work has been funded by the Polish Ministry of Science and Higher Education subvention within the task “Cognitive and sustainable communication systems” in 2020.
The simulations have been based on the QCM simulator from Huawei Technologies Sweden Research Center.






\bibliographystyle{elsarticle-num-names}
\bibliography{sample.bib}







\end{document}